\documentclass{article}

\PassOptionsToPackage{numbers, compress}{natbib}

\usepackage[preprint]{neurips_2021}

\usepackage[T1]{fontenc}    
\usepackage[colorlinks=false, pagebackref=false]{hyperref}       
\usepackage{url}            
\usepackage{booktabs}       
\usepackage{amsfonts}       
\usepackage{nicefrac}       
\usepackage{microtype}      
\usepackage{xcolor}         
\usepackage{graphicx}         
\usepackage{amsmath}
\usepackage{color, colortbl}
\definecolor{Gray}{gray}{0.9}

\usepackage{bbm}

\title{CALM: Contrastive Aligned Audio-Language Multirate and Multimodal Representations}

\author{%
  Vin Sachidananda$\thanks{Work done while at Apple.}$, Shao-Yen Tseng$^*$, Erik Marchi, Sachin Kajarekar, Panayiotis Georgiou
  \thanks{Corresponding author} \\
  \texttt{ vsachi@stanford.edu, \{shaoyent, emarchi, skajarekar, georgiou\}@apple.com} \\
   Apple \\
}

\begin{document}

\maketitle

\begin{abstract}
Deriving multimodal representations of audio and lexical inputs is a central problem in Natural Language Understanding (NLU). In this paper, we present Contrastive Aligned Audio-Language Multirate and Multimodal Representations (CALM), an approach for learning multimodal representations using contrastive and multirate information inherent in audio and lexical inputs. The proposed model aligns acoustic and lexical information in the input embedding space of a pretrained language-only contextual embedding model. By aligning audio representations to pretrained language representations and utilizing contrastive information between acoustic inputs, CALM is able to bootstrap audio embedding competitive with existing audio representation models in only a few hours of training time. Operationally, audio spectrograms are processed using linearized patches through a Spectral Transformer (SpecTran) which is trained using a Contrastive Audio-Language Pretraining objective to align audio and language from similar queries. Subsequently, the derived acoustic and lexical tokens representations are input into a multimodal transformer to incorporate utterance level context and derive the proposed CALM representations. We show that these pretrained embeddings can subsequently be used in multimodal supervised tasks and demonstrate the benefits of the proposed pretraining steps in terms of the alignment of the two embedding spaces and the multirate nature of the pretraining. Our system shows 10-25\% improvement over existing emotion recognition systems including state-of-the-art three-modality systems under various evaluation objectives.
\end{abstract}

\section{Introduction}
Unsupervised and self-supervised representation learning are critical tools in a wide range of machine learning tasks in \emph{natural language processing} (NLP), \emph{natural language understanding} (NLU), computer vision, and speech and audio processing. In particular, these approaches learn feature representations from large amounts of input data, without explicit supervision, which are then utilized in downstream tasks often with small amounts of labeled data. In NLP, context has been employed in deriving embeddings such as word2vec \cite{Mikolov2013Efficient-Estim} or GloVe \cite{Pennington2014GloVe:-Global-V} to produce real valued representations of words from large amounts of unlabeled text. Furthermore, these concepts have been extended to sequence to sequence models \cite{Sutskever2014Sequence-to-seq} in order to encode whole sentences and allowed integration of higher level context. Recently, bidirectional contextual embeddings, such as BERT \cite{Devlin2019BERT:-Pre-train}, have been introduced which are able to incorporate more general forms of context dependent on a particular input sequence through the use of compositions of multi-head self-attention. In this paper, we propose Contrastive Aligned Audio-Language Multirate and Multimodal Representations (CALM), an approach for learning contextual representations of both audio and language modalities in a shared representation space. We find that CALM is able to make use of contrastive and multirate information intrinsic to spoken language inputs and efficiently extends pretrained language models to provide performant audio-language representations. 

Contextual and contrastive prediction has been utilized significantly in both NLP and, more recently, in acoustic problem settings. Early efforts include employing context to learn behavioral embeddings \cite{Li2017Unsupervised-La}, speaker embeddings \cite{Jati2018Neural-Predicti,Milde2018Unspeech:-Unsup} and audio events \cite{Jansen2018Unsupervised-le}. More recent efforts in capturing higher order context, such as in the case of full-length utterances, include the use of more complex tasks and contrastive structures for tasks such as speech recognition \cite{Baevski2020wav2vec-2.0:-A-}.

Due to the nature of spoken language, audio inputs convey both lexical and paralinguistic information, the latter of which can provide meaningful information in tasks such as emotion recognition, intent detection, speaker identification and diarization [Not sure we need to, but can find if we do... add citations]. Along this direction, there have been efforts to augment NLP tasks with this additional information from the acoustic channel. In some instances, bypassing the speech recognition step can lead to NLU models operating end-to-end with audio \cite{Serdyuk2018Towards-end-to-}. In other cases, while the acoustics are not employed towards the actual NLU task they may be used to resolve ambiguity inherent during speech recognition \cite{Shivakumar2018Confusion2Vec:-, Shivakumar2019Spoken-Language,Jeon2020Acoustic-Neighb}. 

More relevant approaches to our work aim at holistically and jointly modeling both acoustic and lexical information. For the sake of conciseness, we discuss those approaches closer to the tasks presented in this paper such as affect, emotions, and behaviors. In \cite{Chung2020SPLAT:-Speech-L}, the authors present a speech-language joint pretraining framework that is used to learn joint representations for NLU tasks. Additionally, contextual models have been used for sentiment analysis \cite{Dos-Santos2014Deep-convolutio, Severyn2015Twitter-sentime}, emotion recognition and behavior annotation \cite{Tseng2017Approaching-Hum,Tseng2019Unsupervised-on}, intent detection \cite{Chung2020SPLAT:-Speech-L}, and improved representations in the case of ambiguity \cite{Shivakumar2018Confusion2Vec:-}.

\section{Overview of contributions\label{sec:overview}}
Our contributions are briefly described here.
\begin{itemize}
    \item Our development employs the notions of short-term stationarity (context) and independence (contrastive) based on multimodal and temporal cues. This allows for low bandwidth streams (e.g. lexical) to be abstracted from coarser temporal context such as utterances rather than subwords or words. This is the fundamental assumption behind this work (Sec. \ref{sec:sts}).
    \item SpecTran: Employs a patch-based transformer on the spectrum (or cepstrum) space in order to create embeddings for small frames of an audio input resulting in ``tokens of audio''. To the best of our knowledge this is a novel contribution in the speech domain (Sec.~\ref{sec:spectran}).
    \item CALP: Pretrains audio embeddings by aligning them in the embedding space with pretrained lexical embeddings. This provides a novel representation that is partly shared among the two modalities, efficient to train, and novel in the audio-language domain (Sec.~\ref{sec:calp}).
    \item We introduce a form of unsupervised learning using a composition of Masked-Language-Modeling (MLM) and Masked-Audio-Modeling (MAM) losses. This approach incorporates multiple modalities in a single transformer model (Sec.~\ref{sec:mlm}).
    \item The embeddings can be employed in various tasks through supervised training of small networks on top of the multimodal embeddings (Sec.~\ref{sec:supervised_heads}). Additionally, a single CALM model is able to operate on unimodal inputs, either audio-only or language-only, in addition to joint audio-language inputs.
\end{itemize}
We show that through our approach, we can achieve substantial gains, especially in the case of the hard-to-label tasks such as emotion recognition. CALM pretraining can also aid in robustness and scalability of pretrained systems. While the experimentation in this work is focused on emotion recognition tasks, we intend to investigate the efficacy of our approach on different tasks, datasets, signal resolutions, and modalities in future work.
Section \ref{sec:contributions} presents more details of the contributions and reasoning for the proposed architecture.

\section{Related Work}

Related work can be characterized into two main directions $\textbf{(I)}$ literature related to the fields of cross-modal pretraining and acoustic-embedding extraction and $\textbf{(II)}$ work in the application domain of emotion recognition.

\subsection{Acoustic embeddings and cross-modal pretraining}
Various models have been employed to compress acoustic information into embedding representations. In unsupervised learning from spectral representations there have been efforts in employing CNN structures \cite{Jati2018Neural-Predicti}, and ResNet models \cite{Jansen2018Unsupervised-le} using triplet networks.
Other efforts employed supervised training for specific tasks. For example, \cite{Ravanelli2018Speaker-recogni} employed a CNN SincNet kernel structure to operate directly on the waveform. \citet{Lin2020An-Efficient-Te} employed an LSTM architecture on the spectral representation. 
\cite{Kowtha2020Detecting-Emoti} employed an LSTM network and a time convolution LSTM network. \cite{Khare2020Self-Supervised} has employed frame-stacking to derive a direct representation for the keys and queries of the cross-modal transformer. Recently \cite{Zhao2021Combining-a-par} employed a CNN architecture with a deep residual network (and a CTC loss).
Yet other approaches have taken an agnostic learning method to derive, using a SincNet convolutional framework, multiple existing knowledge based descriptors \cite{Pascual2019Learning-Proble}.
Indirectly related to our work but important to the pretraining task is the effort by \cite{Baevski2020wav2vec-2.0:-A-} that employs CNN networks of very short duration (25\,ms) audio segments.
There is also a plethora of autoencoder-like systems for pretraining in the audio modality, e.g. \cite{Deng2014Autoencoder-bas} with various architectures. Recent work in vision transformers \cite{Dosovitskiy2020An-Image-is-Wor}, which encode and reconstruct linear patches of images using multi-head attention, is most similar to our architecture for learning representations for audio frames.

Also in cross-modal learning, there have been multiple efforts in the speech domain, although significantly related works are the vision-language cross-modal training frameworks, such as for captioning based on image content.  For emotions and behavior the audio (A), vision (V), and language (L) modalities are often used, however most efforts focus on single modal or two-modality (A/L) systems. 
Some examples with three modalities include \cite{Liang2018Computational-m} that employed a simple yet effective dynamic fusion graph between the A/V/L modalities. 
In \cite{Khare2020Self-Supervised} a three modality setup is obtained via two transformers that share input text as a query field with separate keys and values for A/V modalities.
\citet{Tzirakis2021End-to-end-mult} have investigated a range of fusion techniques, including concatenation, hierarchical attention, self-attention, residual self-attention, and cross-modal hierarchical self-attention. In all of these cases, the systems were supervised and learned the two modalities concurrently.

In our case, we want to exploit existing pretrained systems in the lexical modality to learn in the acoustic modality. Some work along this direction includes \cite{Tseng2021Multimodal-Embe} where an ELMo network is used to jointly train the two modalities and \cite{Siriwardhana2020Jointly-Fine-Tu} where a BERT-like self-supervised architecture is employed. \citet{Radford2021Learning-transf} has aligned lexical and visual embeddings by applying a contrastive objective to match images to textual captions. This, similar to our work, assumes a dependence in the two modalities and also similar to our work employs different representation dimensions.

\subsection{Multi-modal emotion recognition}
Towards our goal of cross modal and multirate pretraining, we selected as a first task for evaluation the task of emotion recognition. This was to benefit from the significant inter-dependency between the modalities inherent in human expressiveness. For example, in expressing emotions speakers will likely express the same emotional content in both modalities, e.g. ``I am upset'' may sound acoustically distressed.

 There has been significant work in emotion recognition in recent years. Much of that work dealt with corpora that are collected in a lab or controlled setting and are thus not going to be examined in this work, for example IEMOCAP \cite{Busso2008IEMOCAP:-Intera} which was collected in clean recording conditions of acted scenarios with few speakers. In our case we want to evaluate in more realistic data so we evaluate on the CMU-MOSEI and UTD MSP Podcast datasets.
There are many works in emotion recognition on these data including on emotional primitives, such as valence, activation, and dominance \cite{psyemo},  and categorical emotions \cite{Ekman1999Basic-emotions}. On emotional primitives  \cite{Lin2020An-Efficient-Te} employed a technique on MSP whereby the feature size of each utterance remained fix via changing the window overlap. This novel method may have some drawbacks in a real-time scenario of unknown word-chunks but nevertheless performs well. Further to that \cite{Kowtha2020Detecting-Emoti} has employed the MSP corpus in addition to proprietary data for the same task.

In our work we focus mostly on categorical emotion recognition. In this task the best performing systems in literature on CMU-MOSEI to the best of our knowledge are by \cite{Khare2020Self-Supervised} and \cite{Mittal2020M3er:-Multiplic} that employed all three modalities on the CMU-MOSEI dataset. The best two-modality system was an ELMo architecture employing only the lexical and acoustic modalities \cite{Tseng2021Multimodal-Embe}. Recently  \cite{Delbrouck2020A-Transformer-b} described a transformer-based system and based on authors' code we achieved slightly better performance at 66.5\% weighted accuracy. This matches our independent and concurrent work in transformers that employed a CNN front end and a multimodal BERT model which achieved 66.6\%.

On the MSP dataset, \cite{Lotfian2018Predicting-cate}  proposed a multitask learning system to jointly model primary and secondary emotions. Importantly they analyze the human performance (via the inter-annotator agreement) and provide an estimated human-performance metric. Prior work has also shown that machine-learning systems can improve over the average annotator \cite{Tseng2019Unsupervised-on} and in some such cases alternative evaluation metrics have to be established.

\section{CALM Architecture \label{sec:contributions}}

\subsection{Short-term stationarity and contrastive elements\label{sec:sts}}
Our work below assumes that short term stationarity holds for the information of interest, that is, nearby audio frames will very likely encode the same target information.
To give a few specific examples, it is more likely that nearby frames of audio are generated by the same speaker and likely contain the same behavioral content (i.e. speaker unlikely to change or their emotions to drastically change), as established by \cite{Li2017Unsupervised-La}. Similarly for the active  speaker \cite{Jati2018Neural-Predicti} or for the audio events \cite{Jansen2017Large-Scale-Aud}. This assumption has to be revisited when the information of interest changes, e.g. frames/ms in the case of phoneme modeling \cite{Baevski2020wav2vec-2.0:-A-} versus seconds for  speaker identification, emotion, or  behavior recognition. 
In many tasks, annotation happens at a coarse scale because of this reason. e.g. \cite{Burmania2015Increasing-the-} employs segments at 2.75 to 11 seconds to annotate emotions; 
\cite{Chakravarthula2021An-analysis-of-} presents an analysis of lexical observation requirements for behaviors where original annotation was on 10 minute scale \cite{Christensen2004Traditional-ver}; and speaker ID is often defined for segments of 4 to 20 seconds.  \cite{Chung2018Voxceleb2:-Deep}.

Similarly we can  assume that this stationarity holds across modalities as well. We can think of the two modalities as being encoded through the same generator, that being the speaker (brain, articulation, acoustic environment).  
Thus there are several ways that this assumption can manifest in practice: speakers will tend to have specific articulations for specific words, which creates an inter-modality dependence; or emotions and behavioral states will affect the choice of words and vice versa.  
Sometimes these can even be dependent on the environment of the speaker; something often undesired and removed via augmentation, e.g.  \cite{Chung2018Voxceleb2:-Deep}. 
For example, talking over the car speakers while driving may change the articulation of specific words to reflect the noisy and far field environment.
These assumptions may be weakened, yet not eliminated, by external factors, such as for example if the choice of words is somehow dictated by another task, e.g. reading or speaking with an assistant agent, that restrict the free choice of vocabulary.

\subsection{Spectral transformer\label{sec:spectran}}
\begin{figure}[t]
    \centering
    \includegraphics[width=0.7\linewidth]{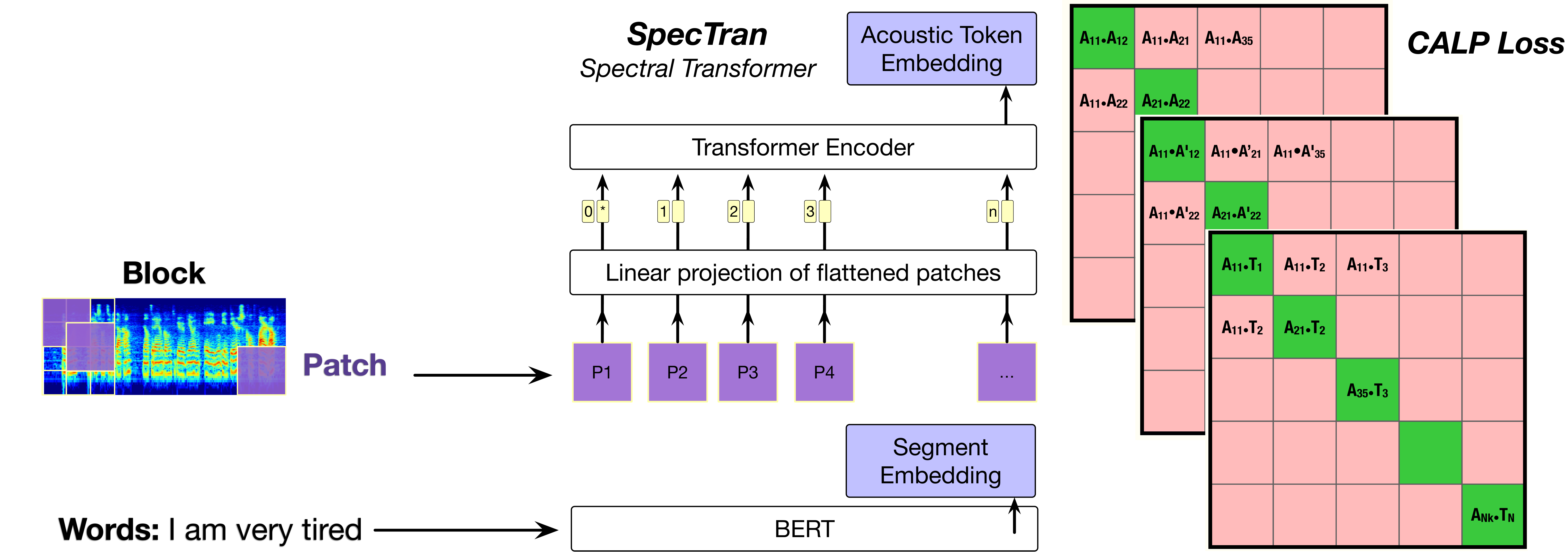}
    \caption{The figure shows the first step for training CALM. Patches of audio are linearized and passed in the transformer network to train SpecTran. SpecTran is expected to capture correlations between short term spectral patches and longer term lexical information via the CALP loss.}
    \label{fig:SpecTrans}
\end{figure}

For encoding audio frames we utilize a Spectral Transformer (SpecTran) whose architecture follows the work in \cite{Dosovitskiy2020An-Image-is-Wor}. The spectral block is converted into a set of linearized patches. The patches can be of arbitrary size and overlap, covering the whole acoustic block. Additionally, a positional embedding is computed for each patch which is incremented first across time and then across frequency band. Those are fed into a transformer network and the first output embedding is then used downstream in supervised heads as shown in Fig. \ref{fig:SpecTrans}.

\subsection{Contrastive Acoustic Language Pretraining: Single block audio with coarse language information\label{sec:calp}}
Based on the short-term stationarity assumption we propose a \emph{Contrastive Acoustic Language Pretraining (CALP)} step, which is based on the efficient multimodal contrastive learning approach in \cite{Radford2021Learning-transf}. In this step, we allow nearby data to generate similar embeddings. 
We assume a recording of sufficient context for the construct of interest; in our case here we are investigating emotions, affect, speaker ID so above a few words is usually sufficient.

Let's denote such a group of $N$ audio files as $A_i$, where $i=[0,N]$. In each of these audio files, we can choose an \emph{acoustic token} to be of fixed length, e.g. 1 second long audio, and represent that as $A_{ij}$, where $j$ is an index into the audio. For example assuming an audio token of 1 second with no overlap then $A_{4,5}$ will correspond to the 5th second in the 4th file.

We denote the corresponding language as $T_i$, since language is a lower bandwidth signal we can choose the language blocks to be longer. For example, someone can say ``I'm feeling so tired today I didn't sleep last night'' and we will need significant amount of language to detect their fatigued state, but that may be captured from just a short sample of audio. Similarly with speaker ID a lot more language will be needed to capture speaker-specific word choices versus listening to their voices.

In most cases we have utterance-level segmentation of audio (if not we can automate this based on pauses) so here we  assume that we use all the language from the utterance (since lower rate information) while only employing a fixed length audio. We thus drop the temporal index from the language.
This can create similar pairs $A_{ij}$ with $A_{i(j+1)}$ and $A_{ij}$ with $T_{i}$ while it can create contrastive pairs like $A_{ij}$ with $A_{kl}$ and $A_{ij}$ with $T_{m}$ where $i\neq k$ and $i \neq m$.
\begin{equation}
\begin{array}{lccccl}
    \text{Close in embedding space: \ } &   A_{ij} & \simeq & A_{i(j+1)} & \simeq & T_{i}  \\
    \text{Far in embedding space: \ } &    A_{ij} & \neq   & A_{kl}     &  \neq   & T_{m}  \text{ where }  i\neq k \text{ and }  i \neq m.
  \end{array}
\end{equation}
Note that in the case of augmentation in any modality the corresponding augmented pairs can also be employed (e.g.  $A_{ij}$ with $A'_{i(j+1)}$)

Given the shuffled nature of a minibatch, we ensure that we capture one set of $A_{ij}$, $A_{i(j+1)}$, $T_{i}$ from each utterance $i$. We then construct the loss matrix $M\times M$ for the minibatch of size $M$. The optimization objective is to minimize elements on the diagonal (same audio-file) while off-diagonal elements are pushed apart. A visualization is shown on the right of  Fig. \ref{fig:SpecTrans}.

A weighted composite NTXent contrastive loss \cite{Sohn2016Improved-deep-m}, $\mathcal{L}_{CALP, \tau} (A^{t}) = \mathcal{L}_{NTXent}(A^{t}, A^{t+1}) + \alpha \mathcal{L}_{NTXent}(A^{t}, T)$, is optimized to seed coarse audio embeddings by minimizing distances between audio frame and language representations within an utterance and maximizing distances of representations belonging to different utterances. During experimentation, we fix  fixed $\alpha=0.25$; the objective is provided below for a single minibatch of size $M$ with $\tau$ being the standard temperature scaling parameter:

\begin{equation}
\begin{aligned}
\mathcal{L}_{CALP, \tau} (A^{t}_i) =& -\log \frac{\exp(\text{sim}(A^{t}_i, A^{t+1}_i)/{\tau})}{\sum_{\substack{j=1 \\ j \neq i}}^M \exp(\text{sim}(A^{t}_i, A^{t+1}_j)/{\tau})} -\alpha \log \frac{\exp(\text{sim}(A^{t}_i, T_i)/{\tau})}{\sum_{\substack{j=1 \\ j \neq i}}^M  \exp(\text{sim}(A^{t}_i, T_j)/{\tau})}  \\
\mathcal{L}_{CALP, \tau} (A^{t}_i) \leq &  \frac{1}{\tau}\bigg[-\alpha \text{sim}(A^{t}_i, T_i) - \text{sim}(A^{t}_i, A^{t+1}_i) + \max_{j \neq i}\text{sim}(A^{t}_i, A^{t+1}_j) \\ &
+ \alpha \max_{k \neq i}\text{sim}(A^{t}_i, T_k)\bigg] + 2\log(M)
\end{aligned}
\end{equation}

Using Log-Sum-Exp properties, we can see that the objective seeks to maximize the cosine similarity of representations of contiguous frames and a frame and its language representation through the terms $- \text{sim}(A^{t}_i, T_i) $ and $ - \text{sim}(A^{t}_i, A^{t+1}_i)$. Additionally, cosine similarity between an audio frame and negative audio frame and language samples within the minibatch are penalized through the terms $\max_{j \neq i}\text{sim}(A^{t}_i, A^{t+1}_j) $ and $ \max_{k \neq i}\text{sim}(A^{t}_i, T_k)$. 

\subsection{Multimodal Transformer}
\begin{figure}[t]
    \centering
    \includegraphics[width=0.87\linewidth]{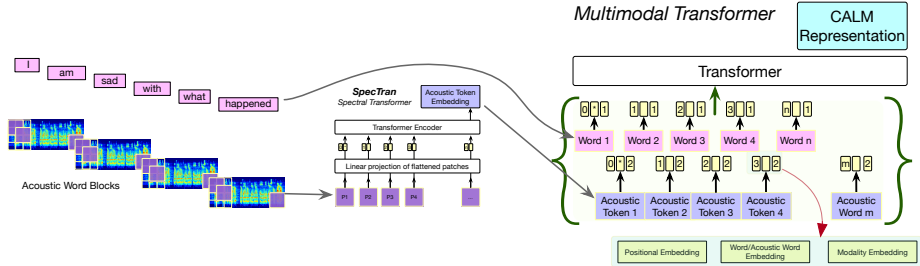}
    \caption{The acoustic tokens, as output by the SpecTran network pretrained via CALP, are employed together with (text) words in a multimodal transformer. The multimodal transformer encodes the set of tokens with modality and positional information.}
    \label{fig:CALM-representation}
\end{figure}

The output of the SpecTran, trained through CALP, is employed in the subsequent multimodal transformer as depicted in Fig.~\ref{fig:CALM-representation}. The set of tokens provided to the multimodal transformer include ``Acoustic Tokens'', learned using SpecTran, and ``Lexical Tokens'', equivalent to the tokens of a pretrained language model. These inputs are accompanied by two embeddings: (i) a positional embedding that corresponds to time in the acoustic modality and to the token index in the language sequence and (ii) a modality embedding. These embeddings are utilized the same manner as the positional and sequence embeddings in the original BERT model \cite{Devlin2019BERT:-Pre-train} where the different forms of input embeddings are combined additively with a subsequent Layer Norm operation. The multimodal transformer is initialized to have weights equivalent to the pre-trained language model used when training SpecTran. In all of our experimentation, we utilize either the BERT base \cite{Devlin2018Bert:-Pre-train} or BERT tiny \cite{turc2019wellread} and denote the resulting CALM models when using each of these pretrained language models as CALM$_{BASE}$ and CALM$_{TINY}$. Language and Audio token sequences are padded separately and inputted to the model in succession which allows for cross-attention between modalities within each multi-head attention block. 

\subsection{Audio-Language Masked Prediction\label{sec:mlm}}

During pretraining, we utilize a masked language modeling (MLM) head for reconstructing masked language tokens and a masked acoustic modeling (MAM) head for reconstructing masked acoustic frames. Both of these masked prediction heads utilize the same architecture as the BERT masked language modeling head, a two layer fully connected network with a GeLU hidden activation.

In constructing our loss function, denote the input audio representations from SpecTran and language tokens for an individual sample as $\{A_{0}, A_{1}, \dots , A_{n}\}$ and $\{T_{0}, T_{1}, \dots , T_{m}\}$ respectively. Additionally, denote the decoding of the Multimodal Transformer outputs using the aforementioned MLM and MAM heads as $\{\hat{A}_{0}, \hat{A}_{1}, \dots , \hat{A}_{n}\}$ and $\{\hat{T}_{0}, \hat{T}_{1}, \dots , \hat{T}_{m}\}$. Note that the decoded acoustic outputs $\hat{A}_i \in \mathbb{R}^{xy}$ where $x$ and $y$ are the dimensions of the input log mel spectogram, used in SpecTran, for the acoustic frame and that the decoded language outputs $\hat{T}_i \in \mathbb{R}^{|V|}$ where $|V|$ is the total number of tokens in the Language Model tokenizer.

During training a subset of input audio frames and language tokens is masked; in the case of language tokens a special [MASK] token replaces $15\%$ of the input tokens while $10\%$ of audio representations and the representations from the subsequent 2 frames are set to the zero vector. Masking audio inputs chunks is necessary to avoid due to the overlap of nearby audio inputs and smoothness of the audio signal.

During training, we minimize the sum of masked audio and masked language modeling losses. For given sequences of corresponding audio, let the sets $K$ and $L$ constitute the audio and language indices being masked. For each masked audio frame, the corresponding loss value is the mean squared error between the original Log Mel Spectogram frame and the predicted output of the MAM head. For each masked language token, the loss is computed as the cross entropy $H(\cdot, \cdot)$ between the predicted token distribution and the true one-hot token distribution. 
\begin{equation}
\begin{aligned}
    \mathcal{L}_{MAM}(\hat{A}^K, A^K) =& \frac{1}{|K|} \sum_{i \in K} (\hat{A}_{i} - A_{i})^2 \\
    \mathcal{L}_{MLM}(\hat{T}^L, T^L) =& \frac{1}{|L|} \sum_{j \in L} H(\hat{T}_{j} |T_j) \\
    \mathcal{L}_{ALT}(A, T) =& \mathcal{L}_{MLM}(\hat{T}^L, T^{L}) + \mathcal{L}_{MAM}(\hat{A}^K, A^{K}) \\
\end{aligned}
\end{equation}

\subsection{Supervised training heads\label{sec:supervised_heads}}

Supervised training can take place on top of the multimodal embeddings of CALM. 
There are two approaches on how to employ the pretrained embeddings.
In the case of frozen pretrained embeddings, multiple heads (shallow feed forward networks) can be added to CALM to achieve multitask learning without any cross-task influences and is the preferred method.
However in some cases we may have tasks and datasets that can benefit from larger amounts of supervised data, in which case we can unfreeze pretraining and allow the CALM network to be adapted to the task.

\section{Experimentation}
\label{experimentation}
We conduct experimentation on two multimodal datasets for emotion recognition:  CMU MOSEI  \cite{Zadeh2018Multi-attention} and  UTD MSP-Podcasts \cite{Lotfian2017Building-natura}. 
We include a number of baselines from recent literature against which to compare the downstream performance of CALM. Baselines used for comparison either use the same or more data for pretraining compared to CALM.

\subsection{Terminology}
To keep explanations clear we use \emph{frame} to describe a slice of the spectral plane. We use an \emph{Acoustic Token} or \emph{block} for brevity to describe a short window of spectral features that will generate an \emph{acoustic token embedding}, e.g. 1 second of audio or 100x64 (assuming standard 64 dimensional filterbanks or MFCC's and 10ms shift). Within that block we employ  \emph{patches}. These can be of arbitrary size with arbitrary 2-D shift, e.g. 20x20 with 10x16 shift. Note that patches are accompanied by a positional embedding and hence any arbitrary configuration trades-off the size of the patch versus the number of patches.
Multiple acoustic tokens can form an utterance, and acoustic tokens can have an overlap themselves, e.g. 1 second blocks with 0.25 second shift. 

In this work, for consistency, we keep patch size at 10x16 with a stride of (5, 8), and the audio block at (50x64) with 30 frames shift (i.e. 64 dimensional filterbanks, 0.5 seconds block with 0.2s overlap).

\subsection{CMU Multimodal Opinion Sentiment and Emotion Intensity (MOSEI)}

\begin{table}[!t]
\centering \includegraphics[height=36ex]{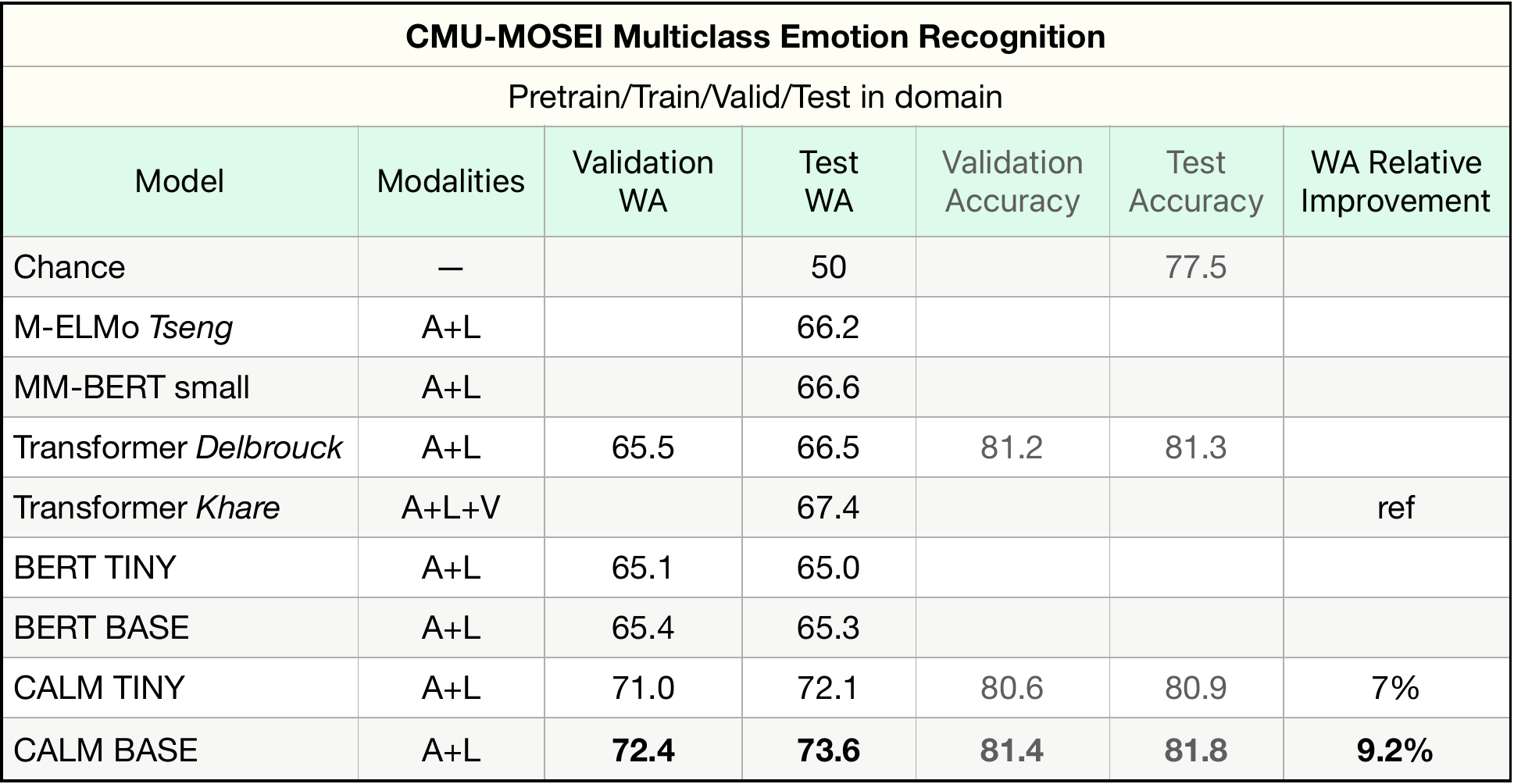}
\caption{CMU-MOSEI emotion recognition results. For completeness in comparing with literature we employ both weighted accuracy and accuracy. 
Weighed accuracy is a more meaningful measure due to the class imbalance between positive and negative labels. 
Comparisons with \cite{Tseng2021Multimodal-Embe}, \cite{Khare2020Self-Supervised} and \cite{Delbrouck2020A-Transformer-b} are shown in table.
Both CALM models show improvement when compared to previous approaches.
Note that for all experiments standard deviation between runs is below 0.41 for weighted accuracy and below 0.58 for accuracy; results from 5 runs with different seed.
\label{tab:cmu}}
\end{table}

The CMU-MOSEI \cite{Zadeh2018Multi-attention} dataset was constructed from YouTube videos featuring more than 1000 speakers and contains 23.5k utterances with Audio, Visual, and Language modalities. In this paper, we utilize only the Audio and Language modalities in the CALM model.
Each utterance in the dataset is annotated for Ekman emotions \cite{Ekman1999Basic-emotions}  of \{happiness, sadness, anger, fear, disgust, surprise\} on a [0,3] Likert scale for presence of emotion. Following previous literature \cite{Zadeh2018Multimodal-lang}, we binarize the labels such that an emotion is said to be present in an utterance for a nonzero Likert score and not present for a score of 0. As multiple emotions can be present in an utterance, this task constitutes a multi-label multi-class classification problem. 

For audio features we use 64 dimensional log spectrum with 10ms shift. For language, to be consistent with literature, we employed the corpus-provided transcripts and removed casing. During fine tuning, we train a supervised head on CALM comprised of a 2 layer MLP, with hidden size of 64, and output predictions for each of the 6 classes. The supervised head operates on top of the [CLS] label.
A binary cross entropy loss is minimized. Both weighted and unweighted accuracy over the validation and test sets are reported in Table \ref{tab:cmu} for CALM and previous approaches. The Adam optimizer \cite{Kingma2014Adam:-A-method-} with a learning rate of $5e^{-5}$ and batch size of 64 is used for training over 50 epochs.

To evaluate the benefits of the various pretraining steps we performed an ablation study as shown in Table \ref{tab:cmu_ablation}. We can see that pretraining helps in performance gains over supervised training. We also see very small gains in performance through incorporating out-of-domain (OOD) data, in this case the MSP podcast data used in the next section. Likely due to the nature of our OOD data the gains were very small. We will employ larger datasets for OOD ablation in the future.

\begin{table}[!t]
\centering \includegraphics[height=33ex]{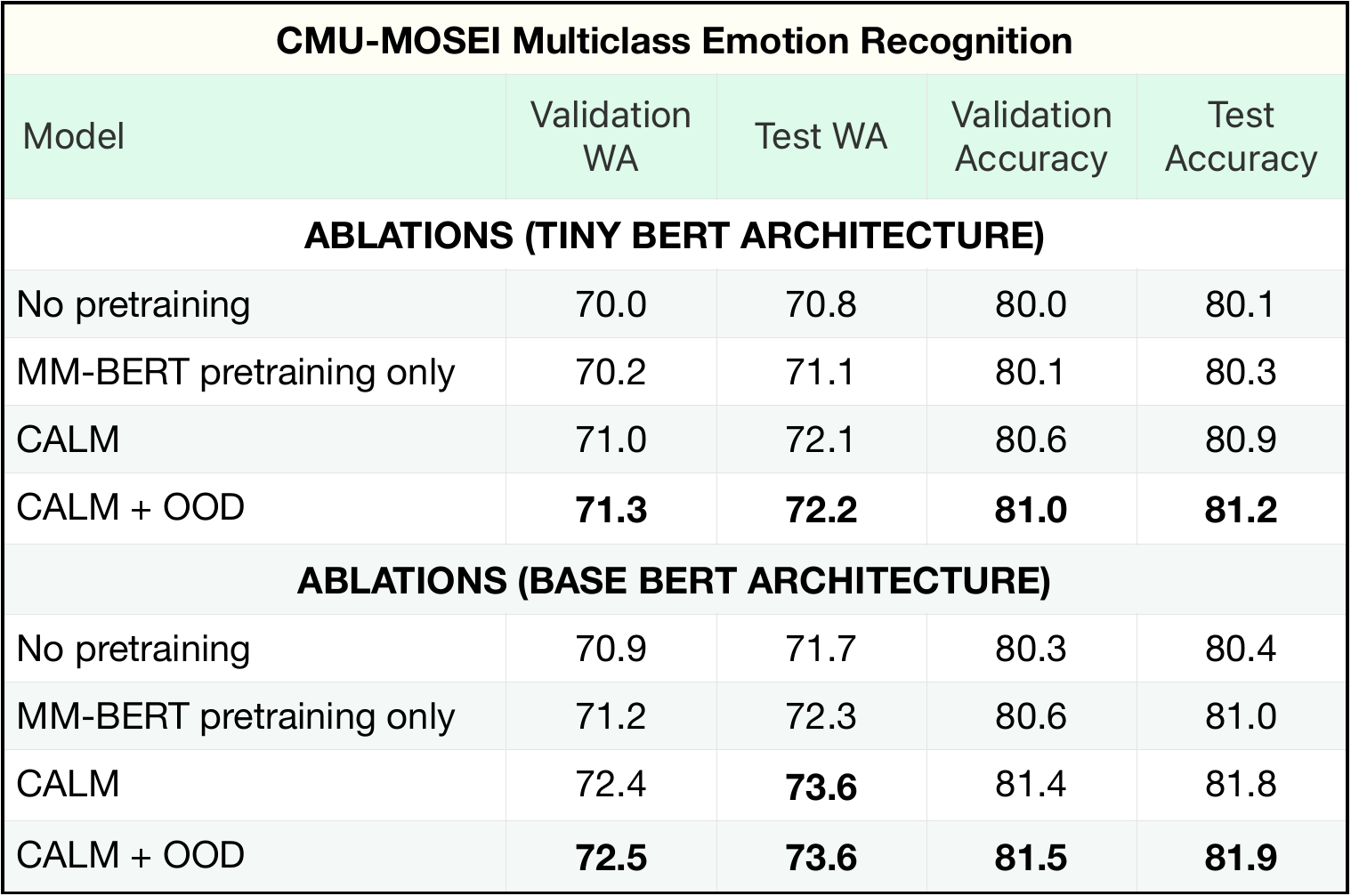}
\caption{Ablation study on CMU-MOSEI. Rows in each section represent: Supervised training only, pretraining only the multimodal transformer, full CALM pipeline, full CALM pipeline with pretraining including MSP Podcast data.
Results show improvement from pretraining multimodal transformer and from training full pipeline. 
\label{tab:cmu_ablation}}
\end{table}

\subsection{UTD Multimodal Signal Processing Podcasts (MSP)}

The UTD MSP-Podcast corpus v1.6 contains about  83 hours of spoken language collected from podcast recordings and about  50k utterances. 
Each utterance is annotated for emotions (Angry,Sad, Happy, Surprise, Fear, Disgust, Contempt, Neutral, Other, No agreement) \cite{Lotfian2017Building-natura}. Annotators are also allowed to choose a secondary emotion. We convert this list into a Prominent emotion (the main emotion annotators chose) and a list of all emotions (Primary + Secondary). This results in a similar setup to CMU-MOSEI and to the one in \cite{Lotfian2018Predicting-cate}.

For audio features we use 64 dimensional log spectrum with 10ms shift. For language we automatically transcribed the data with an ASR system and removed casing. Following previous literature \cite{Lotfian2018Predicting-cate}, we evaluate CALM in predicting emotions as a regression task. During the supervised stage we train a head for the 8 emotions (ignoring as per convention Other and No Agreement) comprised of a 2 layer MLP, with hidden layer of size 64, that outputs a binary label for the 8 categorical emotions using a binary cross-entropy loss. The Adam optimizer  \cite{Kingma2014Adam:-A-method-} with a learning rate of $1e^{-4}$ and batch size of 128 is used for training over 20 epochs. Both weighted accuracy and $F_1$ over the validation and test sets are reported in Table \ref{tab:utd} for CALM and previous approaches.

Note that there are many different evaluation numbers in literature which were difficult to compare with (e.g. using only subset emotion classes and rejecting the remaining). It is also difficult to find papers employing automated (ASR generated) transcription and to employ the lexical modality. Further the dataset is evolving and different versions are employed by different authors. Nevertheless we see a big improvement in both F1 and accuracies from the comparable baselines in literature.

\begin{table}[th]
\centering \includegraphics[height=22ex]{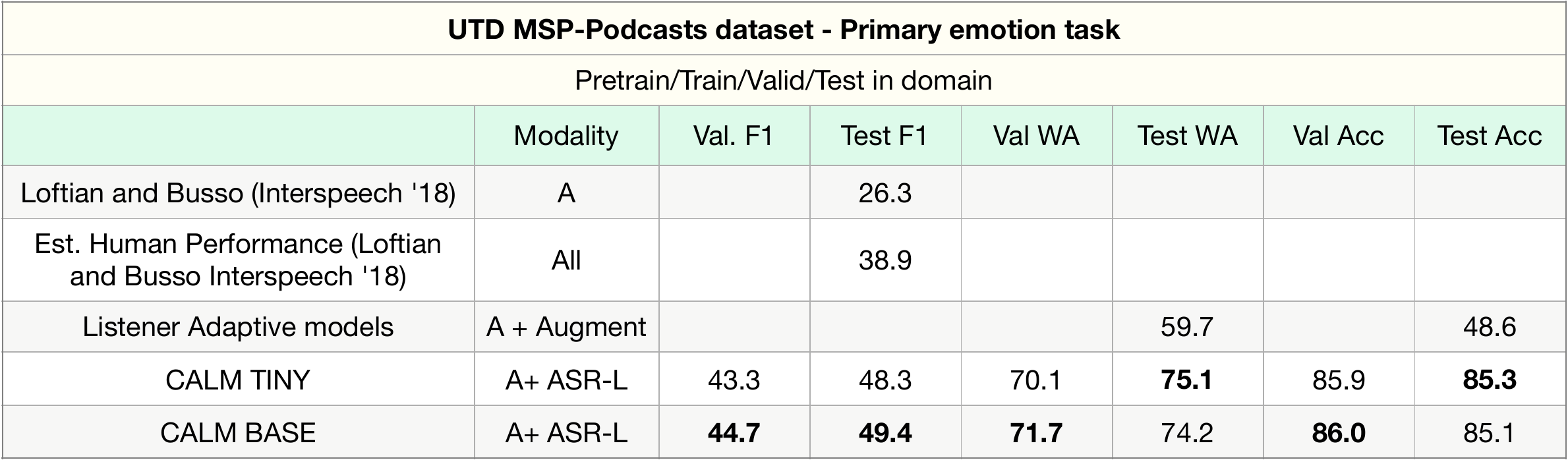}
\caption{Results on the UTD MSP-Podcasts corpus. Comparison points with \cite{Lotfian2018Predicting-cate} and \cite{Ando2021Speech-Emotion-}. 
Note that we were unable to find good A+L comparison points in literature, likely due to most teams not using automatically generated transcripts due to their high word error rate. Nevertheless our approach is robust to such errors and we use the ASR-Language representation. We do not claim that is the best comparison point but the one available.
Further we note the estimated performance from  \cite{Lotfian2018Predicting-cate} may reflect annotator disagreements due to the variability of the data. Our system is trained on the aggregate annotator opinions so it performs more inline with the average human annotator which may explain the much better F1 scores.
Nevertheless, results show improvements over existing data points with both TINY and BASE models.\label{tab:utd}}
\end{table}

\section{Discussion}

In this section, we review the performance of the CALM model on downstream spoken language understanding tasks and the computational efficiency of training CALM. 

\subsection{Performance}
Based on experimental results, we find that CALM provides performance improvements relative to baselines across different metrics and datasets in emotion recognition. We saw an improvement on both CMU-MOSEI and MSP-Podcasts datasets. We also saw that the pretraining was important in providing about 2\% absolute improvement in WA on the CMU-MOSEI task. The gains were roughly equisplit through the CALP pretraining and the MM-BERT pretraining, thus demonstrating the importance of both tasks.  Our ablations also showed that even BERT$_{\text{TINY}}$, with its much smaller parameter space, can provide good gains.
 We believe that introducing large amounts of varied data in pretraining will not only improve performance but will lead to increase robustness to channel and speaker characteristics.

\subsection{Computational and Resource Efficiency}

The computational and resource efficiency of CALM is dependent on three factors: (i) whether a pretrained language model is utilized, (ii) the size of the language model to which the audio modality is aligned, and (iii) whether external audio data, i.e. outside of a dataset's training set, is used for pretraining. When utilizing an open-source pretrained language model, CALM is efficient when compared to other multimodal representation learning approaches as minimal training is performed for the lexical modality. Additionally, the method is aligning of audio representations to the pretrained language model thus exploiting an existing representation space.
In our experiments, CALM pretrains joint audio-language representations on both the CMU-MOSEI and MSP-Podcasts datasets in less than 3 hours on 8 Nvidia Tesla V100 GPUs. 

In this paper, in addition to BERT$_{\text{BASE}}$, we also evaluate CALM in highly resource constrained settings by using a compressed pretrained language model, BERT$_{\text{TINY}}$, and performing experimentation in the setting where only the training set is used for pretraining. Despite it's reduced parameter space the CALM$_{\text{TINY}}$ representation still outperforms other SOTA algorithms. In Table \ref{tab:comp_resource} below, the parameter sizes and training times for CALM, when used with different pretrained language models and dataset sizes, are detailed. 

\begin{table}[th]
\centering \includegraphics[height=13ex]{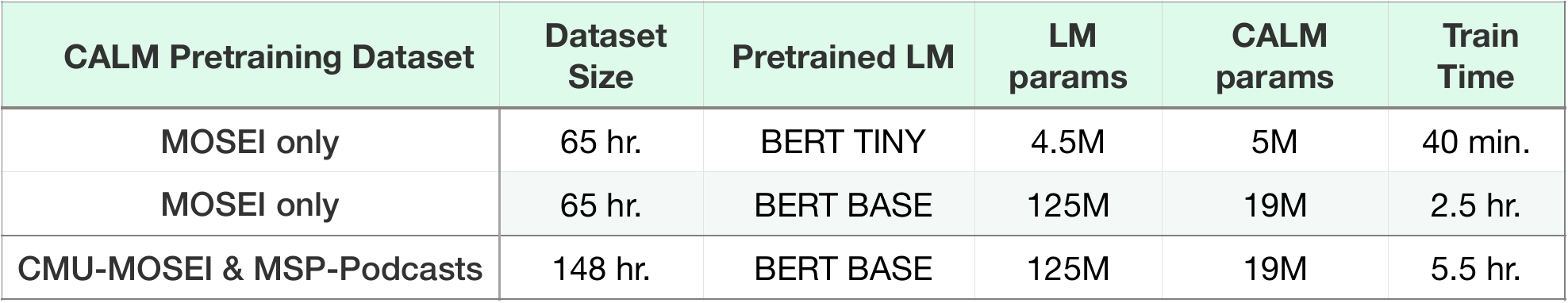}
\caption{Compute and time utilization by pretraining corpus setting}
\label{tab:comp_resource}
\end{table}

\section{Conclusion}

We introduced CALM, a pretraining framework for learning multimodal audio-language representations aligned in a common input space, such as that of a pretrained language model. CALM is flexible in its resource requirements, both due to its ability to leverage pretrained language models and learn audio representations from small amounts of supervised training data. Additionally, the two components of CALM, a multimodal contrastive learning framework and an individual audio-language transformer, are novel in the context of multimodal speech processing. We evaluate on emotion recognition as a downstream task and show that CALM outperforms previous approaches. CALM provides a novel, efficient approach to learning joint audio-language embeddings in a common and multimodally aligned representation space. 

\bibliographystyle{panos}
\bibliography{ref.bib}

\clearpage

\clearpage
\appendix


\section{Appendix}

\subsection{Importance of Audio-Language embedding alignment}
One of the contributions of CALM is the CALP pretraining step that aligns audio with text representations. In order to identify the contribution of this pretraining step to overall performance, we perform ablations by removing all lexical pretraining in CALP, while preserving the audio-to-audio pretraining and all supervision. This is equivalent to setting $\alpha=0$ in equation 2 and removing lexical modality from the multimodal transformer.  We see from the tables below that although we are not employing the lexical modality in inference, incorporating the audio-language alignment step in CALP provides performance gains through cross-modal information transfer.
\begin{table}[th]
    \centering
    \includegraphics[width=0.7\linewidth]{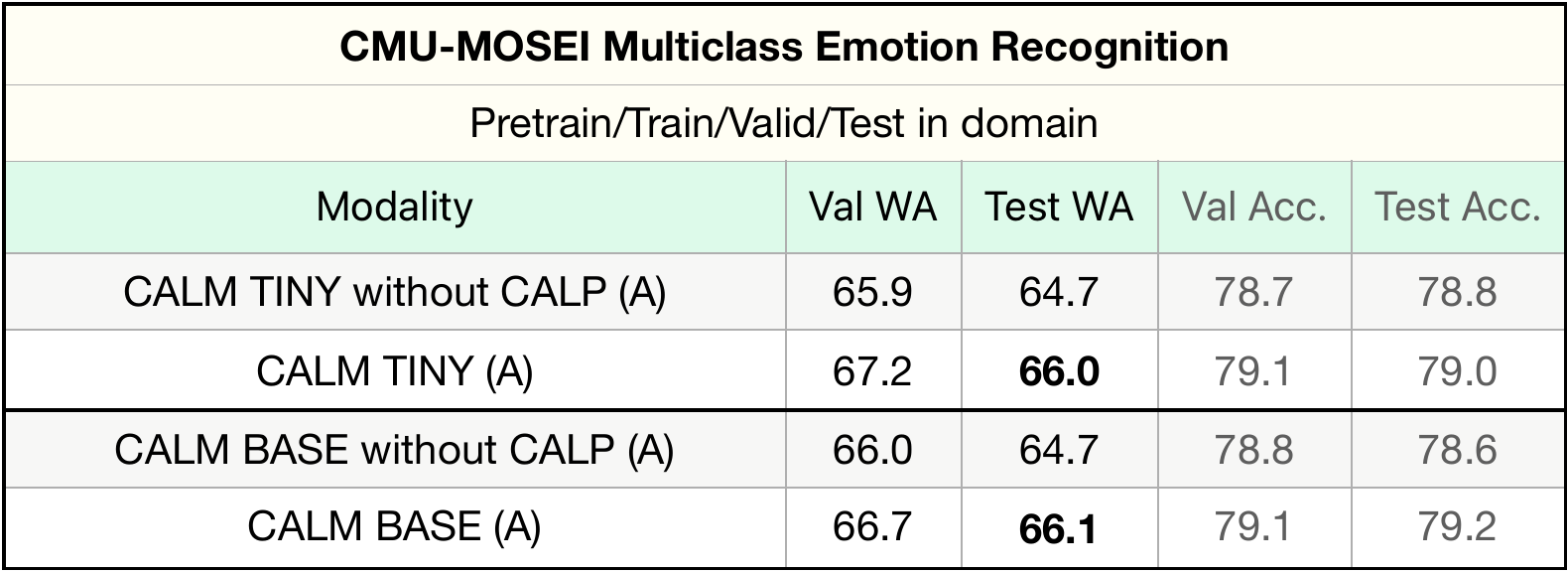}
    \caption{Two training conditions differ only in CALP step containing the audio-lexical embedding alignment. We see  gains from this in both BASE and TINY models}
    \label{tab:calp-cmu}
\end{table}

\begin{table}[th]
    \centering
    \includegraphics[width=0.85\linewidth]{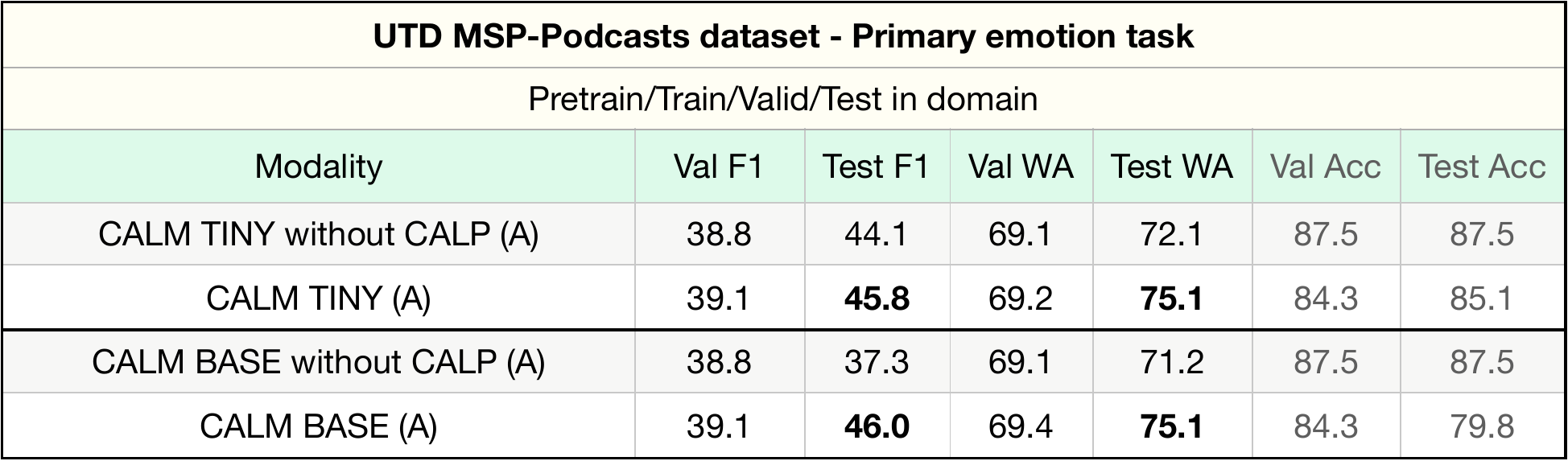}
    \caption{Two training conditions differ only in CALP step containing the audio-lexical embedding alignment. We see gains from this in both BASE and TINY models. Gains in the BASE case are larger.}
    \label{tab:calp-utd}
\end{table}

As seen in Table \ref{tab:calp-cmu} and \ref{tab:calp-utd}, both CMU and MSP datasets have gains from the embedding alignment. We see a larger improvement in the BASE case for MSP-Podcasts and we reason this can be explained by the tougher lexical conditions of this dataset (due to ASR transcription) and better lexical to audio knowledge transfer from the BASE model.

The performance improvements are consistent and strongly support the notion of the modality alignment of CALP. This infers that we can also use lexical information for pretraining even if during test time lexical transcriptions are not available.

\clearpage
\subsection{Modality Ablation}
As expected we see that both modalities perform very well but most of the gains can be achieved with a bimodal system. 
\begin{table}[th]
    \centering
    \includegraphics[width=0.7\linewidth]{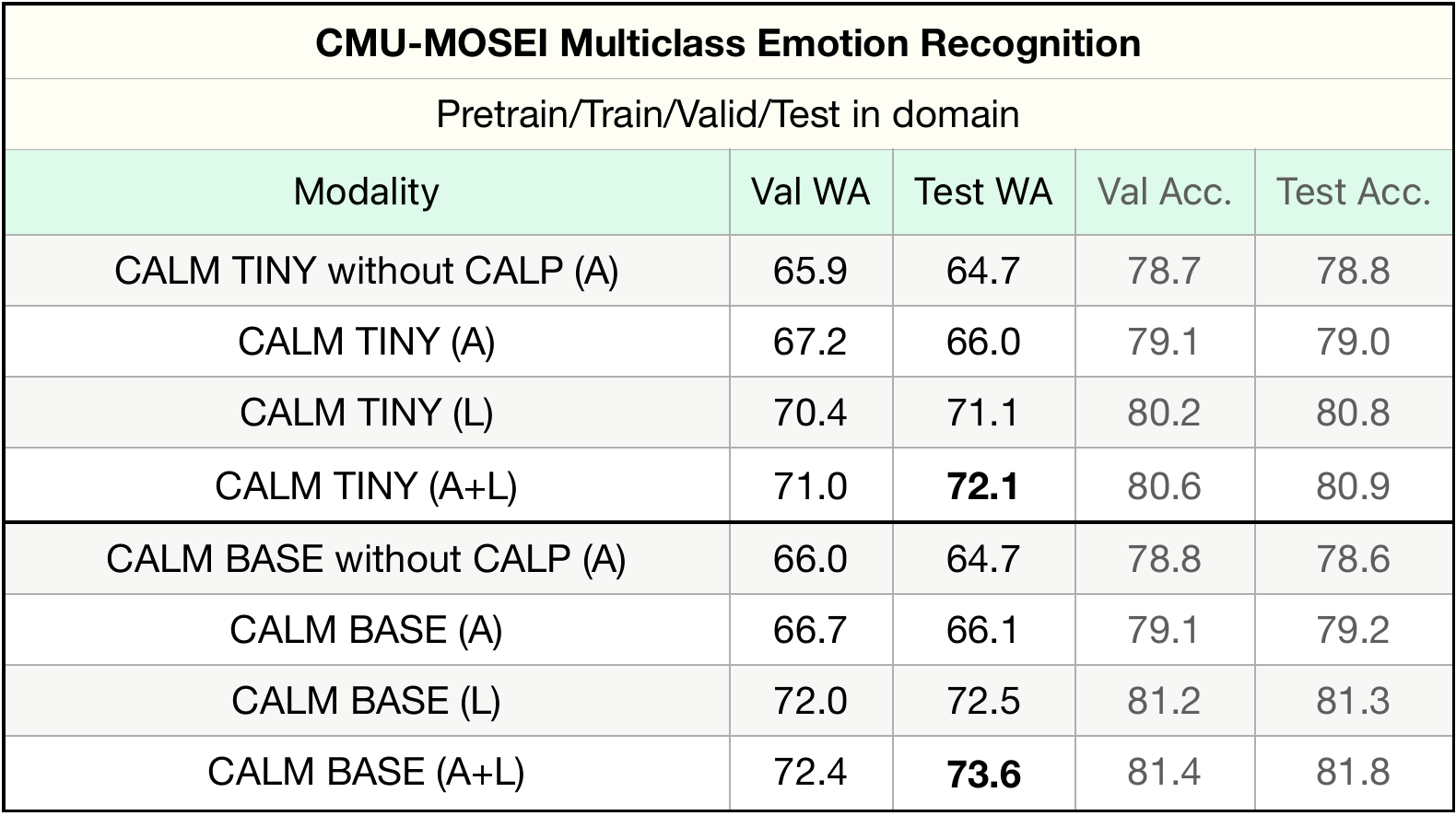}
    \caption{CMU modality ablation}
    \label{tab:modalities-cmu}
\end{table}

\begin{table}[th]
    \centering
    \includegraphics[width=0.85\linewidth]{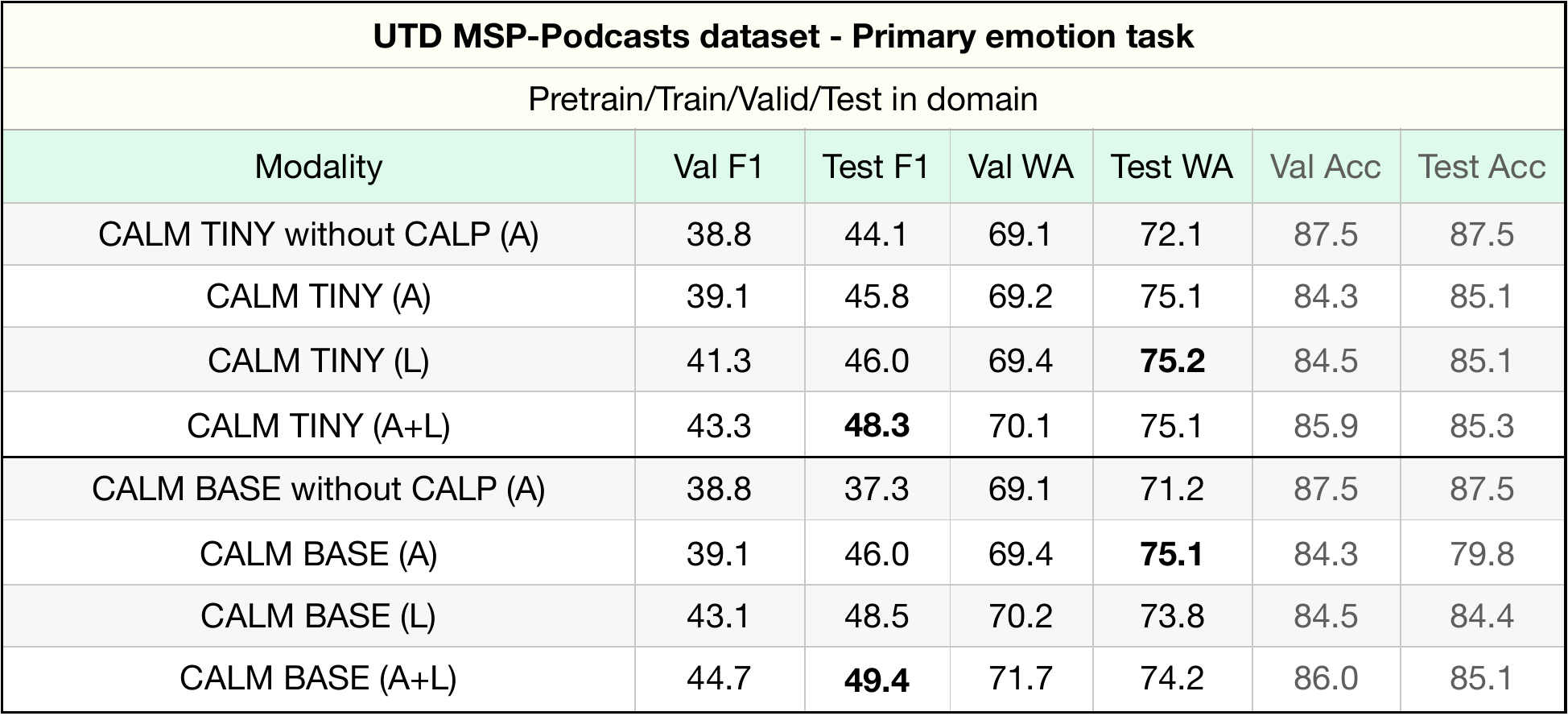}
    \caption{MSP-Podcasts modality ablation. We see that automatically derived transcripts are helpful irrespective of introduced errors in transcription} 
    \label{tab:modalities-utd}
\end{table}

\clearpage
\subsection{Pretraining and limiting supervision}
We wanted to check if pretraining allowed for training with limited supervised data. While as expected having more supervised data provided the best system we also see that limited data also allowed for good performance. This will be a useful aspect in learning new representations in data starved conditions as is often the case for example in complex human behavioral domains.

\begin{table}[th]
    \centering
    \includegraphics[width=0.65\linewidth]{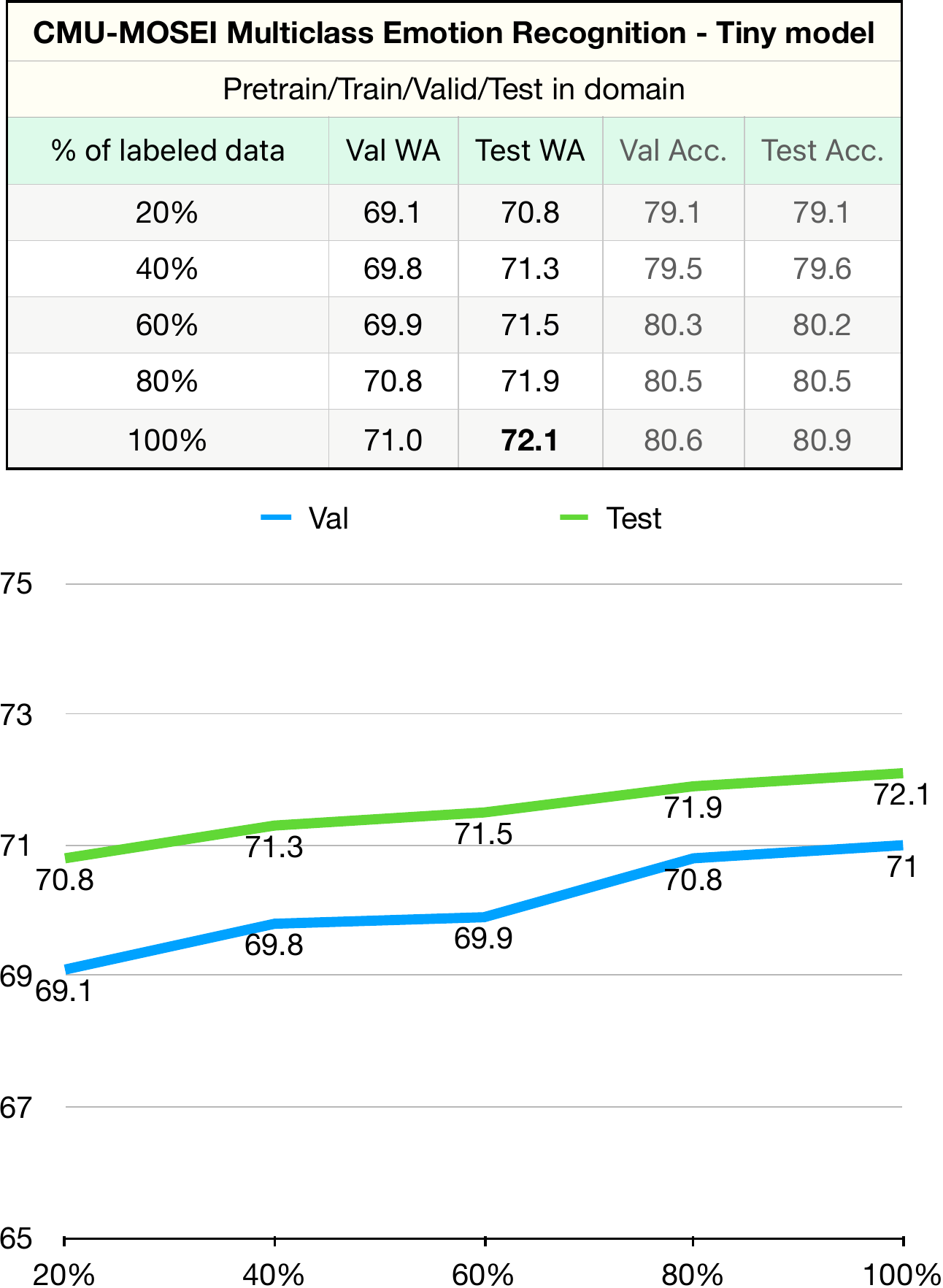}
    \caption{Performance of CALM on CMU with limited supervision data. We see that even with 20\% we outperform existing state-of-the-art systems.}
    \label{tab:piece-cmu}
\end{table}

\begin{table}[th]
    \centering
    \includegraphics[width=0.75\linewidth]{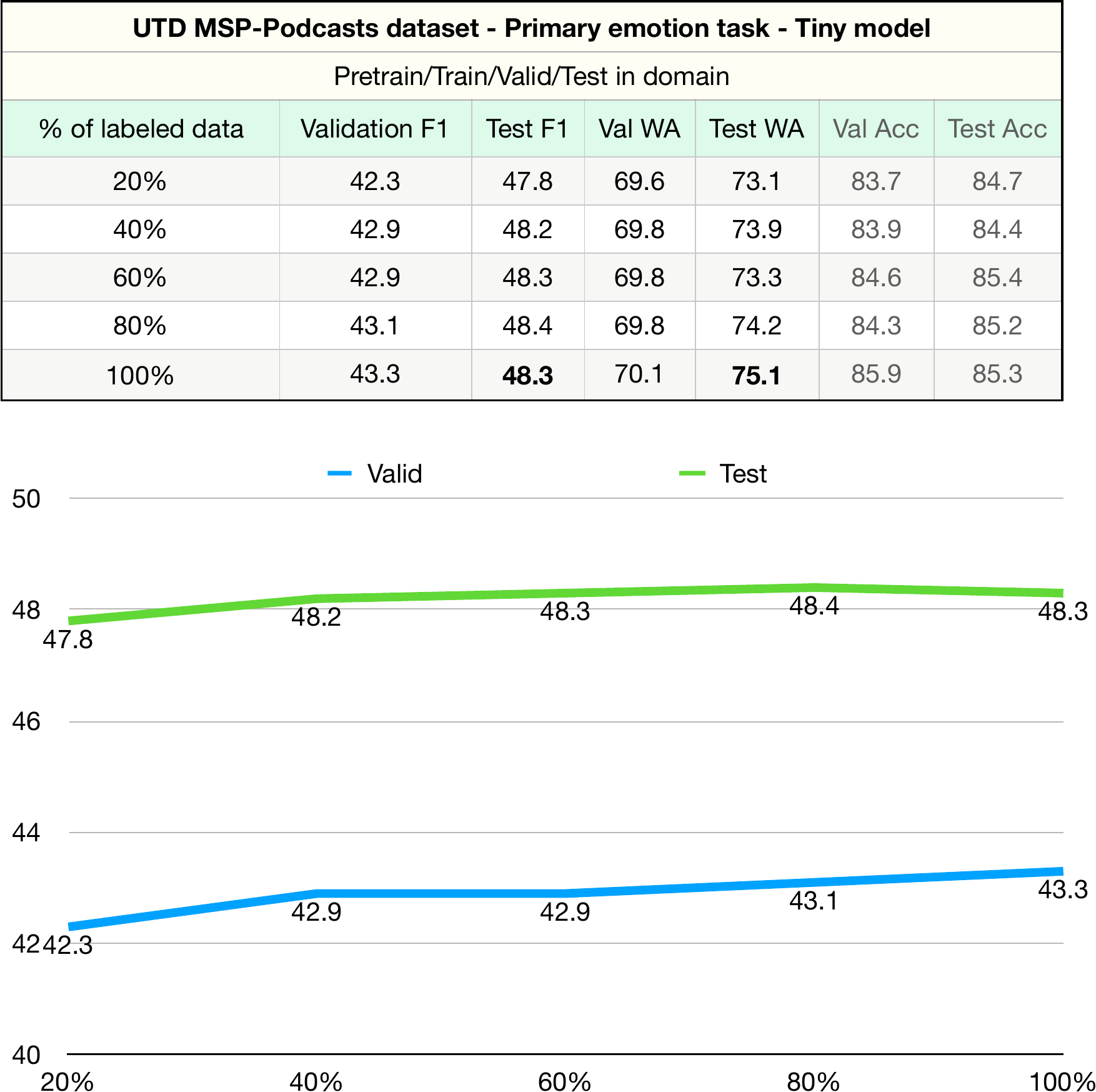}
    \caption{Performance of CALM on MSP-Podcasts with limited supervision data. We see that even with 20\% we outperform existing state-of-the-art systems.}
    \label{tab:piece-utd}
\end{table}

\clearpage
\subsection{Ablation on patch size}
We wanted to see if the SpecTran requires specific configurations in patch size. In internal discussions there were arguments towards patches that cover the whole spectral content. The search space in this case also depends on the acoustic block size. To reduce the parameter space we fixed the block size (as in the whole paper) and changed the patch size and stride.

\begin{table}[th]
    \centering
    \includegraphics[width=0.7\linewidth]{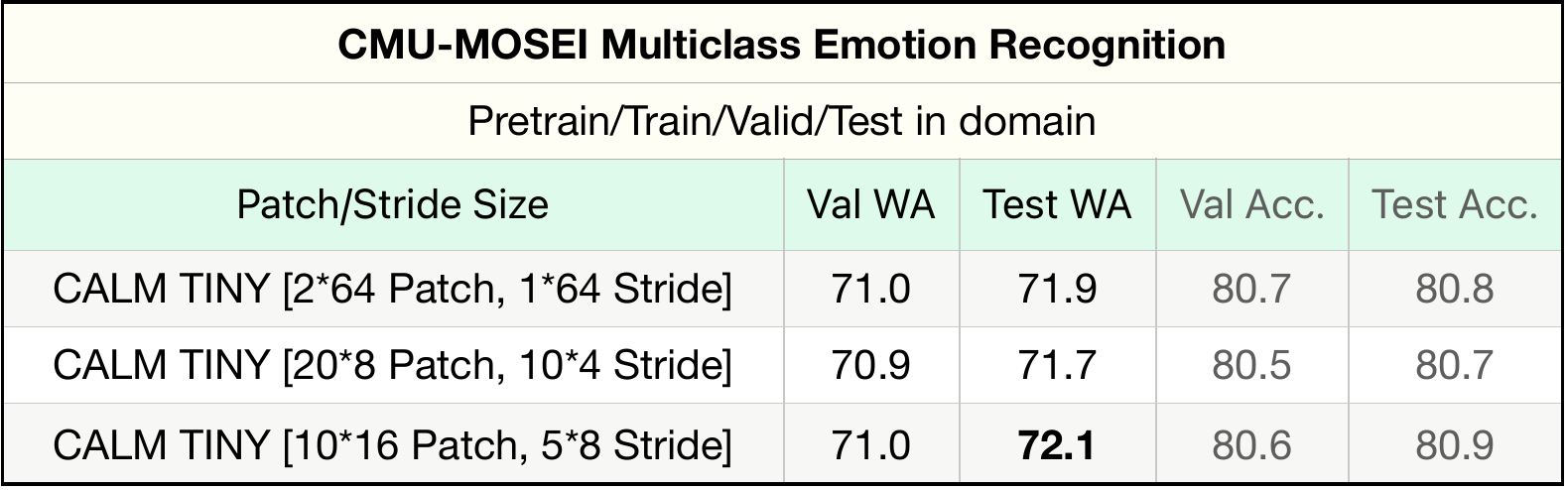}
    \caption{Performance on CMU with different patch sizes}
    \label{tab:patch-cmu}
\end{table}

\begin{table}[th]
    \centering
    \includegraphics[width=0.9\linewidth]{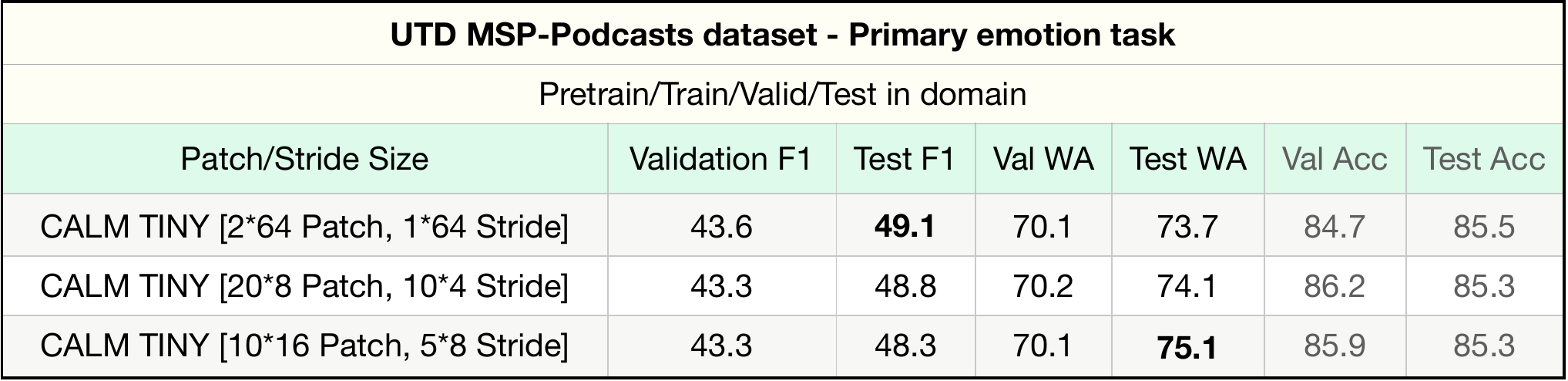}
    \caption{Performance on MSP-Podcasts with different patch sizes}
    \label{tab:patch-utd}
\end{table}

From the experiments above, we notice that the SpecTran network is able to integrate information irrespective of how that is presented in terms of the patches. While this is not an exhaustive search this is a promising indication that the SpecTran system can be robust to such choices.

\section{Evaluation Metrics}
In this section, we provide definitions for the evaluation metrics used during experimentation: Weighted Accuracy (WA), Unweighted Accuracy (Acc), and micro-averaged F$_1$ score. For notation, let $C$ denote the set of output classes, $|C|$ the number of output classes and $\mathcal{TP}, \mathcal{TN}$ the total number of positive and negative labels in the evaluation set.

$$WA = \frac{1}{2|C|} \sum_{c \in C} \frac{TP_c}{P_c} + \frac{TN_c}{N_c}, \hspace{5mm} Acc = \sum_{c \in C} \frac{TP_c + TN_c}{P_c + N_c}$$
$$F_1^{micro} =  \sum_{c \in C} \frac{TP_c + TN_c}{\mathcal{TP} + \mathcal{TN}} \frac{TP_c}{TP_c + 1/2(FP_c + FN_c)}$$

\end{document}